\documentclass[prl,twocolumn,superscriptaddress,showpacs,preprintnumbers,amsmath,amssymb]{revtex4}

\usepackage[english]{babel}

\usepackage{graphicx}
\usepackage{epsfig}
\usepackage[dvips]{color}
\usepackage{float}

\usepackage{amsfonts,amsmath}

\usepackage{hyperref}

\def\iotabar{\lower3pt\hbox{$\mathchar'26$}\mkern-7mu\iota}

\newcommand {\aplt} {\ {\raise-.5ex\hbox{$\buildrel<\over\sim$}}\ }
\newcommand{\dd}{\mbox{d}}

\begin{document}

\title{{
\small
\bf{Fractional generalization of Fick's law: a microscopic approach}}}

\author{I. Calvo}
\affiliation{Laboratorio Nacional de Fusi\'on, Asociaci\'on
EURATOM-CIEMAT, 28040 Madrid, Spain}

\author{R. S\'anchez}
\affiliation{Fusion Energy Division, Oak Ridge National Laboratory,
Oak Ridge, TN 37831, U.S.A.}

\author{B. A. Carreras}
\affiliation{BACV Solutions Inc., Oak Ridge, TN 37830, U.S.A.\vskip2cm}

\author{B. Ph. van Milligen}
\affiliation{Laboratorio Nacional de Fusi\'on, Asociaci\'on
EURATOM-CIEMAT, 28040 Madrid, Spain}

\date{\today}

\begin{abstract}
In the study of transport in inhomogeneous systems it is common to
construct transport equations invoking the inhomogeneous Fick
law. The validity of this approach requires that at least two
ingredients be present in the system. First, finite characteristic
length and time scales associated to the dominant transport
process must exist. Secondly, the transport mechanism must satisfy
a microscopic symmetry: global reversibility. Global reversibility
is often satisfied in nature. However, many complex systems
exhibit a lack of finite characteristic scales. In this Letter we
show how to construct a generalization of the inhomogeneous Fick
law that does not require the existence of characteristic scales
while still satisfying global reversibility.
\end{abstract}

\pacs{05.40.Fb, 05.10.Gg, 05.60.-k}

\maketitle

Over a hundred years ago, A. Fick proposed the famous formula that
bears his name whilst studying transport processes in saline
aqueous solutions~\cite{Fick}. He observed that a salt flux was
driven by the presence of a salt density gradient (in the absence
of other forces) and hypothesized that they were linearly related
to each other. In one dimension, Fick's relation would read:
\begin{equation}\label{eq:Fickhom}
\Gamma_F = - A \frac{\partial n}{\partial x}\ ,
\end{equation}
where $n$ is the salt density and the minus sign makes sure that
the flux tends to reduce the gradient. Since then, Fick's law has
found application in countless systems, and expressions like Eq.
(\ref{eq:Fickhom}) are applied to the transport of many
quantities: electrons and ions, neutral particles, energy or
momentum, risk, chemical reactants, or currency, just to name a
few.

Using Fick's expression in the standard continuity equation yields
the familiar classical diffusive equation,
\begin{equation}\label{eq0}
\frac{\partial n}{\partial t} = A \frac{\partial^2 n}{\partial
x^2}\ ,
\end{equation}
where $n$ represents now the density of the quantity of interest.
The constant $A$ is the diffusive coefficient, which is related to
some well-defined characteristic length ($l$) and time ($\tau$)
scales of the underlying dominant transport process (in the case
of salt transport, the mean-free-path between collisions with the
background water molecules and the inter-collision time): $A\sim
l^2/\tau$. By well-defined, we mean that $l$ and $\tau$ are both
finite and much smaller than the system size and lifespan,
respectively. 

Many physical systems are, however, inhomogeneous. The usual
inhomogeneous extension of Fick's law is:
\begin{equation}\label{eq3}
\Gamma_F = - A(x) \frac{\partial n}{\partial x} ~~~ \Rightarrow
~~~ \frac{\partial n}{\partial t} = \frac{\partial}{\partial
x}\left[A(x) \frac{\partial n}{\partial x}\right].
\end{equation}
In spite of its familiarity, this expression is not the only possible
extension that reduces to Eq. (\ref{eq0}) when
$A(x)=A\equiv\mbox{const.}$ In fact, Fick's law is a particular case
of the more general Fokker-Planck diffusive law~\cite{Kam}, that
defines the flux by,
\begin{equation}\label{eq5}
\Gamma_{\rm FP} =  B(x)n - \frac{\partial}{\partial x}
\left[ A(x) n\right].
\end{equation}
Eq. (\ref{eq3}) requires that $B(x) = \dd A(x)/\dd x$.  Traditionally,
the validity of this constraint (and thus Eq.  (\ref{eq3})) has been
justified by invoking the presence of a particular symmetry of the
underlying microscopic transport mechanism: {\it local reversibility}
(LR)~\cite{Kam}. The symmetry applies when the probability (per unit
time) of a particle being transported from any location $x$ to any
other location $x'$ equals that of being transported from $x'$ to $x$
[For instance, consider reactants moving in a reactive medium. LR
holds whenever the probability rate of moving between any $x$ and $x'$
without reacting depends on the \emph{total number} of reactive
centers found along the reactant path. Similar situations are very
common in nature.]. The condition for the validity of Fick's law is
however less stringent than LR: global reversibility (GR) is
sufficient. Namely, that the probability (per unit time) of a particle
leaving $x$ be the same as the probability (per unit time) of
arriving at $x$. Clearly, LR implies GR but the converse is not
true~\cite{Kam}.

The purpose of this Letter is to revisit this discussion in a
different but related context, which has emerged as an important
paradigm in the last decade: that of scale-free (or fractional)
diffusion~\cite{Metzler00,Zaslavsky02}. Scale-free transport appears
in systems in which some of the transport characteristic scales, on
which Fick's law is fundamentally based, are absent.  The question we
will try to answer in this Letter is: which is the expression that
\emph{correctly} represents fluxes in an \emph{inhomogeneous} system
satisfying GR but lacking such characteristic scales? The relevance
and timeliness of the answer is justified by the recent surge of
interest in physical, economical, biological and social systems in
which scale-free transport is
observed~\cite{Metzler00,Zaslavsky02}. An example is provided by those
systems in which (particle or energy) transport takes place via
correlated avalanches. This seems to be the case of some magnetically
confined fusion plasmas~\cite{Newman96,Carreras96,Sanchez01}, the
propagation of forest fires~\cite{Drossel92},
earthquakes~\cite{Shaw92} or solar flares~\cite{Lu91}. Transport
events in these cases have a maximum size that is only limited by the
system size $L$, and therefore a characteristic size that diverges
with (some power of) $L$. In some of these systems, LR is satisfied at
least in certain limits. For instance, in the propagation of forest
fires, the transition rate at which fire propagates from tree A to
tree B would heavily depend on how many active trees exist between
them, thus satisfying LR.

The discussion will proceed in parallel to the previous one on
classical diffusion. We start by reviewing the scale-free version of
the homogeneous Fick law~\cite{Metzler00,Zaslavsky02}. The relevant
transport equation is then expressed in terms of fractional
differential operators (FDOs)~\cite{Podlubny}. To understand why, we
consider the simplest one, known as the Markovian, symmetric
fractional diffusion equation (MsFDE):
\begin{equation}\label{eq1}
\frac{\partial n}{\partial t} =
A_\alpha\frac{\partial^\alpha n}{\partial |x|^{\alpha}},
~~~ \alpha\in(0,2],
\end{equation}
where $A_\alpha$ is a constant. The Riesz FDO is defined as
\begin{equation}
\frac{\partial^\alpha}{\partial |x|^\alpha} \equiv
\frac{-1}{2\cos(\pi\alpha/2)}\left({}_{-\infty}D^{\alpha}_x +
{}^{\infty}D^{\alpha}_{x}\right),
\end{equation}
where the Riemann-Liouville FDOs of order $\alpha$ are
~\cite{Podlubny}:
\begin{eqnarray}\label{poi}
{}_a D^{\alpha}_x  f &\equiv&
\frac{1}{\Gamma(m-\alpha)} \frac{\dd^m}{\dd x^m}\int_{a}^x \frac{f(x')
}{(x-x')^{\alpha-m+1}}\ \dd x',~~~~\\\label{poi2}
{}^a D^{\alpha}_{x}  f &\equiv&
\frac{(-1)^{m}}{\Gamma(m-\alpha)} \frac{\dd^m}{\dd x^m}\int_x^{a}
\frac{f(x')}{(x'-x)^{\alpha-m+1}}\ \dd x',\nonumber
\end{eqnarray}
being $m$ the integer number verifying that $m-1\leq\alpha<m$. In spite
of their complicated appearance, the theory of fractional differential
operators is very well-established~\cite{Podlubny}. The feature of
FDOs which makes them relevant to the context of scale-free transport
is their action in Fourier space:
\begin{equation}\label{eq:propertyRL}
{\cal F}[{}_{-\infty}D_x^\alpha f]=(-ik)^\alpha\hat f, \quad{\cal
F}[{}^{\infty}D_x^\alpha f]=(ik)^\alpha\hat f,
\end{equation}
where $\hat f(k)\equiv{\cal F}[f](k)$. Using property (\ref{eq:propertyRL})
it follows that the Riesz FDO satisfies:
\begin{equation}
{\cal F}\left[\frac{\partial^\alpha f}{\partial
|x|^\alpha}\right](k)=-|k|^\alpha \hat f(k).
\end{equation}
This result justifies the name ``\emph{fractional differential
operators}'', since they naturally extend the notion of derivative to
non-integer order. What is most relevant to us is that they also
provide a suitable generalization of classical diffusion. Taking the
Fourier transform of Eq. (\ref{eq1}), we obtain:
\begin{equation}
\frac{\partial \hat n(k)}{\partial t} = - A_\alpha |k|^\alpha \hat
n(k),
\end{equation}
which reduces to the classical diffusive equation (Eq.  (\ref{eq0}))
for $\alpha=2$ (note that for $\alpha=2$ the Riesz FDO yields the
second derivative operator). This fractional extension is relevant for
systems in which transport characteristic scales diverge. To
illustrate this point, let us examine the propagator (i.e. the
temporal evolution of an initial condition $n(x,0) = \delta(x-x_0)$)
of Eq. (\ref{eq1}). The time-derivative of the second moment of the
propagator is easily related to the transport characteristic length
scale $l$. In the case of the diffusive equation ($\alpha=2$) the
propagator is a Gaussian function,
$$G(x,t|x_0,0) = (4\pi At)^{-1/2}
\exp\left\{-(x-x_0)^2/(4At)\right\},$$
which has a finite second
moment that increases linearly with time and thus yields a finite
$l$. However, for $\alpha<2$, the propagator is given by a symmetric
L\'evy distribution of order $\alpha$ and behaves like
($|x-x_0|>\hspace{-0.1cm}>1$):
\begin{equation}
G(x,t|x_0,0) \sim C_\alpha t^{-1/\alpha}
\left(\frac{|x-x_0|}{t^{1/\alpha}}\right)^{-(1+\alpha)}\ ,
\end{equation}
where $C_\alpha$ is a constant~\cite{Taqqu}. Due to its fat
power-law tail, the second moment of this propagator is infinite.
Thus, the transport described by this equation lacks a finite
characteristic length. This result, together with the fact that
L\'evy distributions are  stable with respect to the central limit
theorem (as the Gaussian law)~\cite{Taqqu}, endows FDOs with the
physical basis needed to justify their use to model scale-free
transport. Let us define, for convenience, the operator
\begin{equation}\label{eqf}
{}_{a}F_x^{\beta} \equiv
\frac{-1}{2\cos(\pi(\beta+1)/2)}\left({}_{-a}D_x^{\beta} -
{}^{a}D_x^{\beta}\right).
\end{equation}
Then, the fractional particle flux associated to Eq. (\ref{eq1}) can
be written as:
\begin{equation}\label{flux}
\Gamma_{F;\alpha} = -A_\alpha \left({}_{\infty}F_x^{\alpha-1} n\right).
\end{equation}
Evidently, the operator ${}_{a}F_x^{\beta}$ satisfies that:
\begin{equation}
\frac{\partial}{\partial x} \cdot \left( {}_{\infty}F_x^{\alpha -
1}\right) = \frac{\partial^\alpha}{\partial |x|^\alpha}.
\end{equation}
Eq. (\ref{flux}) is the fractional generalization of the homogeneous
Fick law, Eq. (\ref{eq:Fickhom}), which is recovered for $\alpha=2$
due to the fact that ${}_{\infty}F_x^{1} = \dd/\dd x$.

We now come to derive the central result of this Letter: the
inhomogeneous extension of (\ref{eq1}) that preserves GR or, in
other words, the scale-free (fractional) generalization of the
inhomogeneous Fick law, Eq. (\ref{eq3}). We start by giving
precise expressions for the GR/LR symmetries and then we present a
new way to derive the inhomogeneous Fick law (Eq. (\ref{eq3})).
The need for a new approach is caused by the failure of the
standard methods to derive the inhomogeneous Fick law when
characteristic length scales are lacking. Finally, we show that
the new procedure works in the fractional case, yielding the
fractional generalization of the inhomogeneous Fick law. For the
sake of simplicity, we restrict ourselves to the discussion of the
Markovian case, but the extension to include non-Markovianity is
rather straightforward.

The Continuous-Time Random Walk (CTRW) \cite{MonWei} provides a
suitable framework to derive macroscopic transport equations from the
microscoscopic dynamics. CTRWs are models describing a large number of
particles whose motion is defined probabilistically using a
probability density function (pdf). For our purposes, it is enough to
consider one-dimensional separable models defined by two pdfs: a
\emph{step-size} pdf, $p(\Delta_x; x)$, and a \emph{waiting-time} pdf,
$\psi(\Delta_t; x)$, $\Delta_t\geq0$. The joint pdf
$\xi(\Delta_x,\Delta_t; x)=p(\Delta_x; x)\psi(\Delta_t;x)$ gives the
probability that a particle located at $x$ at time $t$ jumps to
$x+\Delta_x$ at time $t+ \Delta_t$. As mentioned above, we restrict
the discussion to Markovian problems by selecting a Poisson law for
the waiting time pdf: $\psi(\Delta_t;x) =
\tau(x)^{-1}\exp\left(-\Delta_t/\tau(x)\right)$, where $\tau(x)$ is
the mean waiting time at $x$. Imposing conservation of the number of
particles leads~\cite{KreMonSch,MilSanCar04} to the Generalized Master
Equation (GME) governing the time evolution of the density of
particles $n(x,t)$:
\begin{eqnarray}\label{eq:GME}
\frac{\partial n(x,t)}{\partial t} =
\int_{-\infty}^{\infty}
\frac{n(x',t)}{\tau(x')} p(x-x'; x')\dd x' - \frac{n(x,t)}{\tau(x)}.
\end{eqnarray}
The GME is the starting point in the derivation of (macroscopic)
fluid transport equations, by going to the fluid limit in which only
long-time, large-distance information is retained. This can be easily
done in Fourier space by taking the limit of small $k$. The Fourier
transform of the GME (\ref{eq:GME}) with respect to $x$ yields
\cite{MilCarSan05}:
\begin{eqnarray}\label{eq:GMEFourier}
\frac{\partial \hat n(k,t)}{\partial t} = \int_{-\infty}^{\infty}
\frac{n(x',t)}{\tau(x')} \left(\hat p(k;x')-1\right) e^{ikx'}\dd
x',
\end{eqnarray}
where $\hat p(k;x')=
\int_{-\infty}^{\infty}p(\Delta;x')e^{ik\Delta}\dd\Delta$ is the
characteristic function of $p(\Delta;x')$. We introduce now the
characteristic exponent, $\Lambda(k;x')$, through $\hat p(k;x')\equiv
\exp\Lambda(k;x')$. The fluid limit of the GME is obtained then by
performing the small $k$ approximation $\hat p(k;x')\approx
1+\Lambda(k;x')$, which turns Eq. (\ref{eq:GMEFourier}) into:
\begin{eqnarray}\label{eq:GMEFourier2}
\frac{\partial \hat n(k,t)}{\partial t} = \int_{-\infty}^{\infty}
\frac{n(x',t)}{\tau(x')}  \Lambda(k;x')  e^{ikx'}\dd x'.
\end{eqnarray}

Firstly, let us show how to obtain the homogeneous classical (Eq.
(\ref{eq0})) and fractional (Eq. (\ref{eq1})) diffusive equations from
Eq. (\ref{eq:GMEFourier2}). We assume that both $\tau(x')$ and
$\Lambda(k;x')$ are independent of $x'$. $\tau$ becomes then the
characteristic transport time scale. The characteristic length scale
is related (if it exists) to the variance $\sigma_2$ of the step-size
pdf, $p(\Delta)$. In the absence of external forces, the central limit
theorem guides us to choose either a Gaussian or a symmetric L\'evy
distribution for $p(\Delta)$~\cite{Taqqu}.  The characteristic
exponent of the Gaussian is $\Lambda(k) = - \sigma_2 k^2$, which
inserted into Eq. (\ref{eq:GMEFourier2}) yields the (Fourier transform
of the) classical diffusive equation (\ref{eq0}) with
$A=\sigma_2/\tau$. The finite variance of the Gaussian is related to a
finite characteristic length scale $l\sim\sqrt{\sigma_2}$. In the case
of a symmetric L\'evy pdf, the characteristic exponent is~\cite{Taqqu}
$\Lambda(k) = - \sigma_\alpha |k|^\alpha$, with $0<\alpha<2$. The
fluid limit of Eq. (\ref{eq:GMEFourier2}) then reduces to (the Fourier
transform of) Eq. (\ref{eq1}). The infinite variance of the L\'evy
pdfs implies that $l$ diverges.

We will now formalize the microscopic symmetries underlying the
inhomogeneous Fick law (Eq. (\ref{eq3})) that we discussed briefly at
the beginning of this Letter. First, we introduce the $1$-particle
transition rate, $T(\Delta;x)=p(\Delta; x)/\tau(x)$.  The LR
condition is then expressed as~\cite{Kam} $T(-\Delta,x+\Delta) =
T(\Delta,x),\ \forall x$. As mentioned, Fick's law only requires the
less stringent GR condition,
\begin{equation}\label{eq:condFick}
\int_{-\infty}^{\infty} \hspace{-0.2cm}
T(-\Delta,x+\Delta)\dd\Delta= \int_{-\infty}^{\infty}
\hspace{-0.2cm} T(\Delta,x)\dd\Delta,\quad \forall x,
\end{equation}
as will be shown in the following.

To derive the inhomogeneous Fick law in the classical (Gaussian)
case, one would typically enforce the symmetry (Eq.
(\ref{eq:condFick})) after Taylor expanding the right-hand side of
Eq. (\ref{eq:GME}) around $x$ and keeping only the terms involving up
to the second moment of $T$. However, a L\'evy pdf of order $\alpha<2$
does not have finite moments of order equal to or greater than
$\alpha$. Hence, we need to develop a procedure to find the fluid
limit equations while imposing GR which does not require an expansion
in moments. The key point is that, after some straightforward
manipulations, the symmetry (\ref{eq:condFick}) can be recast into the
following, equivalent form:
\begin{equation}\label{eq:condFickFourier}
\int_{-\infty}^{\infty}\left[\frac{\hat
p(k,x')-1}{\tau(x')}\right]e^{ikx'}\dd x'=0, \quad \forall k.
\end{equation}
The fluid limit of the symmetry can be expressed in terms of the
characteristic exponent:
\begin{equation}\label{eq:condFickFourierlowk}
\int_{-\infty}^{\infty}\frac{\Lambda(k,x')}{\tau(x')}e^{ikx'}\dd
x'=0,
\end{equation}
which is the small $k$ approximation of
Eq. (\ref{eq:condFickFourier}).

It is instructive to rederive the ordinary, inhomogeneous Fick
law, Eq. (\ref{eq3}), using this formalism. Consider the most
general form of the characteristic exponent of the Gaussian law
allowed by the central limit theorem~\cite{Taqqu}:
\begin{equation}\label{eq:charexpGaussian}
\Lambda(k;x')=ia(x')k-\sigma_2(x')k^2.
\end{equation}
Inserting this expression into (\ref{eq:condFickFourierlowk}) we
immediately find that the following relation between $\bar
a(x')=a(x')/\tau(x')$ and $\bar\sigma_2(x')=\sigma_2(x')/\tau(x')$
must hold:
\begin{equation}\label{eq32}
\widehat{\bar a}(k)= -ik \widehat{\bar\sigma_2}(k),
\end{equation}
or, in real space:
\begin{equation}\label{eq32b}
\bar a(x')=\frac{\dd\bar\sigma_2(x')}{\dd x'}.
\end{equation}
This relation will be recognized as the one invoked earlier to
reduce the full Fokker-Planck law to the inhomogeneous Fick law.
Using Eqs. (\ref{eq:charexpGaussian}) and (\ref{eq32b}) and
Fourier inverting Eq. (\ref{eq:GMEFourier2}), the classical
inhomogeneous Fick law is recovered:
\begin{eqnarray}\label{eq:FickOrdinary}
\frac{\partial n}{\partial t}=-\frac{\partial
\Gamma_F}{\partial x} ,~~~\Gamma_F=-\bar\sigma_2(x)\frac{\partial
n}{\partial x}.
\end{eqnarray}

We can use this very same scheme to derive the fractional version of
the inhomogeneous Fick law in the setup of L\'evy pdfs with
algebraic tails that are stable according to the central limit
theorem~\cite{Taqqu}. The relevant fractional generalization of
Eq. (\ref{eq:charexpGaussian}) is now:
\begin{equation}\label{eq:charexpLevy}
\Lambda(k;x')=ia(x')k-\sigma_\alpha(x') |k|^\alpha,\quad
\alpha\in(0,2),
\end{equation}
with $\sigma_\alpha(x')>0$. Imposing that Eq.  (\ref{eq:charexpLevy})
satisfy again GR (Eq.  (\ref{eq:condFickFourierlowk})) yields
$ik\widehat{\bar a}(k)-|k|^\alpha\widehat{\bar\sigma_\alpha}(k)=0$,
where $\bar a(x')=a(x')/\tau(x')$,
$\bar\sigma_\alpha(x')=\sigma_\alpha(x')/\tau(x')$. The fractional
version of Eq. (\ref{eq32}) thus becomes:
\begin{equation}\label{eq:relfracFourier}
\widehat{\bar a}(k)=
\frac{|k|^\alpha}{ik}\widehat{\bar\sigma_\alpha}(k),
\end{equation}
Using the identity $2\cos(\pi\alpha/2)|k|^\alpha =
(ik)^{\alpha}+(-ik)^{\alpha}$ we can rewrite
Eq. (\ref{eq:relfracFourier}) as
\begin{equation}\label{eq:relfracFourier2}
\widehat{\bar a}(k)=
\frac{(ik)^{\alpha-1}-(-ik)^{\alpha-1}}{2\cos(\pi\alpha/2)}
\widehat{\bar\sigma_\alpha}(k),
\end{equation}
and making use of the properties of FDOs under Fourier transforms
discussed previously:
\begin{equation}\label{eq:fracrelation}
\bar a(x')= {}_{\infty}F_{x'}^{\alpha-1}\ \bar\sigma_\alpha,
\end{equation}
where the operator ${}_{\infty}F_x^{\alpha-1}$ was defined in
Eq. (\ref{eqf}). Eq. (\ref{eq:fracrelation}) is the fractional
generalization of the classical relation (\ref{eq32b}). Using this
result and Eq. (\ref{eq:charexpLevy}) in Eq. (\ref{eq:GMEFourier2}),
and Fourier inverting, we obtain the sought-for fractional
generalization of the inhomogeneous Fick law, $\partial_t
n=-\partial_x \Gamma_{F;\alpha}$, where the fractional Fick flux is
now given by:
\begin{eqnarray}\label{eqfalpha}
\Gamma_{F;\alpha}=  - \left[{}_{\infty}F_x^{\alpha-1}\left(
\bar\sigma_\alpha n\right) - \left(
{}_{\infty}F_x^{\alpha-1}\bar\sigma_\alpha\right)n\right].
\end{eqnarray}

Eq. (\ref{eqfalpha}) is the central result of this Letter. It provides
the keystone on which any description of transport in systems lacking
characteristic length scales but still satisfying GR (or LR) should be
built. As we mentioned, an example could be provided by the
propagation of forest fires across an inhomogeneous distribution of
trees. Another possible application might be the modeling of some
aspects of radial particle transport in turbulent fusion plasmas, such
as those confined in certain tokamak regimes
\cite{Carreras01,Castillo04}. Additional terms might be needed in this
case to account for additional effects associated to radially varying
temperatures and external forcing. Note also that Eq. (\ref{eqfalpha})
is very different from the extension that one would expect from
Eq. (\ref{flux}) in the spirit of Eq.  (\ref{eq3}). Namely,
\begin{equation}\label{flux2}
\Gamma_{F;\alpha} \neq -\bar\sigma_\alpha(x)
\left({}_{\infty}F_x^{\alpha-1} n\right).
\end{equation}
The reason is that fractional derivatives do not satisfy the Leibniz
rule for the derivative of the product of two functions
~\cite{Podlubny}. Hence, Eq. (\ref{flux2}) is an equality only in the
special case $\alpha=2$. Needless to say, other fractional
generalizations (including the one where the flux is defined by the
right-hand side of Eq. (\ref{flux2})) might also be valid under
different assumptions.

Finally, we would like to point out that the scheme laid out in this
Letter also provides the basis for addressing the derivation of the
fractional Fick law in three-dimensional inhomogeneous systems.  The
CTRW construction can be trivially extended to any dimension. However,
the formulation in terms of fractional derivative operators, even in
cases in which rotational invariance reduces the problem to one
effective dimension, requires to tackle a number of non-trivial
technical details, which we plan to study in a future work.

\vspace{0.3cm} \noindent
\emph{Acknowledgements.-} Part of this research was sponsored by
the Laboratory Research and Development Program of the Oak Ridge
National Laboratory, managed by UT-Battelle, LLC, for the US-DOE
under contract number DE-AC05-00OR22725.

\end{document}